\documentclass[aps,prd,superscriptaddress,amsmath,twocolumn,10pt]{revtex4-1}
\usepackage{color,hangcaption,hhline,psfrag,rotating,amssymb}
\usepackage[hang,nooneline]{subfigure}
\usepackage{dcolumn}
\usepackage{verbatim}
\usepackage{bm}

\begin{document}
\title{Prediction of the bottomonium D-wave spectrum from full lattice QCD}

\author{J.~O.~Daldrop}
\thanks{Current address: HISKP, Universit\"{a}t Bonn, 53115 Bonn, Germany}
\affiliation{SUPA, School of Physics and Astronomy, University of Glasgow, Glasgow, G12 8QQ, UK}
\author{C.~T.~H.~Davies}
\email[]{c.davies@physics.gla.ac.uk}
\affiliation{SUPA, School of Physics and Astronomy, University of Glasgow, Glasgow, G12 8QQ, UK}
\author{R.~J.~Dowdall}
\email[]{Rachel.Dowdall@glasgow.ac.uk}
\affiliation{SUPA, School of Physics and Astronomy, University of Glasgow, Glasgow, G12 8QQ, UK}

\collaboration{HPQCD collaboration}
\homepage{http://www.physics.gla.ac.uk/HPQCD}
\noaffiliation

\date{\today}

\begin{abstract}
We calculate the full spectrum of $D$-wave states in the 
$\Upsilon$ system in lattice QCD for the first time, 
using an improved version of NonRelativistic QCD on coarse and 
fine `second generation' gluon 
field configurations from the MILC collaboration  
that include the effect of up, down, strange and charm 
quarks in the sea. Taking the $2S-1S$ splitting to set 
the lattice spacing, we determine the ${}^3D_2-1\overline{S}$ splitting 
to 2.3\%, and find agreement with experiment.  
Our prediction of the fine structure relative to the ${}^3D_2$ 
gives the ${}^3D_3$ 
at 10.181(5) GeV and the ${}^3D_1$ at 10.147(6) GeV. 
We also discuss the overlap of ${}^3D_1$ operators with 
${}^3S_1$ states. 

\end{abstract}

%\pacs{}
%\keywords{}

\maketitle

{\it Introduction.} The spectrum of $b\overline{b}$ states has provided 
a very important testing ground for strong interaction 
physics because of the number of radial and orbital 
excitations that are `gold-plated', i.e. well 
below the threshold for decay to $B$ mesons. 
The recent discovery of the $\eta_b(1S)$~\cite{exptetab} and $h_b(1P)$ 
and $h_b(2P)$ mesons~\cite{bellehb} 
filled in important gaps in the spin-singlet states.
The mass of the $\eta_b$ meson had previously been 
predicted by lattice QCD~\cite{Gray:2005ur} and the $h_b$ meson masses 
were widely expected, and 
found, to be very close to the spin-average of 
their associated spin-triplet states. 

The key missing gold-plated mesons are now the $\Upsilon(1D)$ 
states. These are very difficult to find experimentally although 
the ${}^3D_2$ has been seen in radiative decay from 
the $\Upsilon(3S)$~\cite{Bonvicini:2004yj}. Masses of the $D$-wave states have 
been predicted in potential model calculations 
(see, for example~\cite{Kwong:1988ae, Brambilla:2004wf}), 
but it is hard to quantify the errors 
in these predictions except by 
using different forms for the potentials. 

In lattice QCD the starting point is QCD itself. 
The difficulties with the $D$-wave 
states then stem from the signal to noise 
ratio; the signal 
falls exponentially in lattice time 
with the $D$-wave mass but the noise falls 
with the smaller ground state $S$-wave mass. 
Very large samples of meson correlators 
on full QCD gluon field configurations 
are then needed to obtain a reliable signal. 
Here we give the first results from lattice QCD that 
are able to distinguish the fine structure of $D$-wave 
states. 

{\it Lattice Calculation.} We use `second generation' gluon field configurations 
recently generated by the MILC collaboration~\cite{Bazavov:2010ru}. 
These have a gluon action fully improved through $\alpha_sa^2$~\cite{Hart:2008sq} 
and include the effect of $u$, $d$, $s$ and $c$ quarks in the 
sea using the Highly Improved Staggered Quark 
formalism~\cite{Follana:2006rc}. 
The $u$ and $d$ quarks have the same mass, $m_l$, 
so the configurations are denoted as $n_f=2+1+1$. 
We use three ensembles to give two values of the 
lattice spacing and two values of $m_l$.  
The parameters of the ensembles are given 
in Table~\ref{tab:params}; we label them 
as 3, 4 and 5 from earlier work~\cite{Dowdall:2011wh} in 
which we mapped out the $S$ and $P$-wave bottomonium 
spectrum and determined the lattice spacing from 
the $\Upsilon$ $(2S-1S)$ splitting. 

We calculate $b$ quark propagators on these configurations using 
an improved lattice discretisation of NonRelativistic QCD (NRQCD). 
NRQCD is an expansion in powers of the heavy quark velocity 
and therefore good for $b$ quarks since
$v^2/c^2 \approx 0.1$ inside their bound states. 
The Hamiltonian includes all terms through $\mathcal{O}(v^4)$~\cite{Dowdall:2011wh}:
 \begin{eqnarray}
 aH &=& - \frac{\Delta^{(2)}}{2 am_b} \\
 & & - c_1 \frac{(\Delta^{(2)})^2}{8( am_b)^3}
            + c_2 \frac{i}{8(am_b)^2}\left(\bf{\nabla}\cdot\tilde{\bf{E}}\right. -
\left.\tilde{\bf{E}}\cdot\bf{\nabla}\right) \nonumber \\
& & - c_3 \frac{1}{8(am_b)^2} \bf{\sigma}\cdot\left(\tilde{\bf{\nabla}}\times\tilde{\bf{E}}\right. -
\left.\tilde{\bf{E}}\times\tilde{\bf{\nabla}}\right) \nonumber \\
 & & - c_4 \frac{1}{2 am_b}\,{\bf{\sigma}}\cdot\tilde{\bf{B}}  
  + c_5 \frac{\Delta^{(4)}}{24 am_b} 
  -  c_6 \frac{(\Delta^{(2)})^2}{64(am_b)^2} \nonumber.
\label{eq:H}
\end{eqnarray}
Here $\nabla$ is the symmetric lattice derivative and $\Delta^{(n)}$ 
is the lattice discretization of the continuum $\sum_iD_i^n$. 
$\bf \tilde{E}$ and $\bf \tilde{B}$ are the chromoelectric 
and chromomagnetic fields.
$am_b$ is the bare $b$ quark mass, which 
is tuned by determination on the lattice of the spin-average of 
ground-state $\Upsilon$ and $\eta_b$ meson masses. 
This was done in~\cite{Dowdall:2011wh}
to give the values used here, quoted in Table~\ref{tab:dval}.

The $v^4$ terms in $\delta H$ have coefficients $c_i$ whose values are 
fixed from matching lattice NRQCD to full QCD, either perturbatively 
or nonperturbatively. 
Here we use coefficients for $c_1$, $c_5$ and $c_6$ that 
include $\mathcal{O}(\alpha_s)$ corrections, as described in~\cite{Dowdall:2011wh}. 
The coefficients $c_3$ and $c_4$ of the spin-dependent 
$v^4$ terms have been tuned from a study of the fine structure 
of the $\chi_b(1P)$ states. We find $c_3=1.0$ with an error of 0.1.   
$c_4$ is significantly larger. Here we use $c_4$ = 1.25 on the 
coarse lattices and 1.10 on the fine lattices. These agree within 
0.1 both with the value required to give $P$-wave fine structure 
in agreement with experiment and with the $\mathcal{O}(\alpha_s)$ 
improved result~\cite{Dowdall:2011wh}.  

To make meson correlators for $D$-wave states we use
a quark 
propagator made from either a local or a smeared source 
which has appropriate derivatives 
applied to it to generate a $D$-`wavefunction'. 
This propagator
is then combined with a local propagator  
and the same derivatives and smearings applied 
at the sink to create a $2\times 2$ matrix of correlators for 
each $D$-wave state. The complete set of combinations of 
spin matrices and derivatives needed is given 
in~\cite{Davies:1994mp}. 
Note that the spin-2 and spin-3 representations split into 
irreducible representations of the lattice rotational group 
$\{A_1,A_2,E,T_1,T_2 \}$, 
which must be considered independently since their masses
can differ by discretisation errors.
Very high statistics is required - we have typically 
32,000 correlators for every source operator per ensemble, 
using multiple time sources per configuration. The time sources 
are binned over for analysis.  
 
Bayesian fitting~\cite{gplbayes} is used to extract the spectrum from 
the correlators using fit function:
\begin{equation}
G_{\mathrm{meson}}(n_{sc},n_{sk};t) = \sum_{k=1}^{n_{\rm{exp}}}a(n_{sc},k)a^*(n_{sk},k)e^{-E_kt}.
\label{eq:fit}
\end{equation}
$E_k$ is the energy of the $(k-1)$th radial excitation and $a(n,k)$ label 
the amplitudes depending on source and sink smearing. We fit all the 
$D$-wave states together taking the ${}^3D_2E$ state as the reference 
state, with a prior of width 0.1 on its ground-state 
energy. Relative 
to that we take prior value $0 \pm 40$ MeV on the ground-state energy 
of the other states.  We take priors $0.5\pm 0.5$ GeV on 
radial excitation energies and $0.1 \pm 1.0$ on amplitudes. 
We fit correlators from time $t/a$ = 2 to 12 except for 
the local-local correlators 
which we take from $t/a$ = 9 to 12. 

\begin{table}
\caption{
Details of the MILC gluon field ensembles used in this paper. 
$a$ is the lattice spacing in fm determined from 
the $\Upsilon$ $(2S-1S)$ splitting and $L/a$ and $T/a$ 
give the lattice size.
$am_{l},am_{s}$ and $am_c$ are the sea quark masses in lattice units. 
Ensembles 3 and 4 are denoted ``coarse'' and 5, ``fine.'' 
}
\begin{ruledtabular}
\begin{tabular}{lllllll}
Set & $a$ (fm) & $am_{l}$ & $am_{s}$ & $am_c$ & $L/a \times T/a$ \\
\hline
3 &  0.1219(9)  & 0.0102  & 0.0509 & 0.635 & 24$\times$64 \\
4 &  0.1195(10) & 0.00507 & 0.0507 & 0.628 & 32$\times$64 \\
\hline
5 &  0.0884(6)  & 0.0074  & 0.037  & 0.440 & 32$\times$96 \\
\end{tabular}
\end{ruledtabular}
\label{tab:params}
\end{table}

\begin{figure}
%[t]
\includegraphics[width=0.8\hsize]{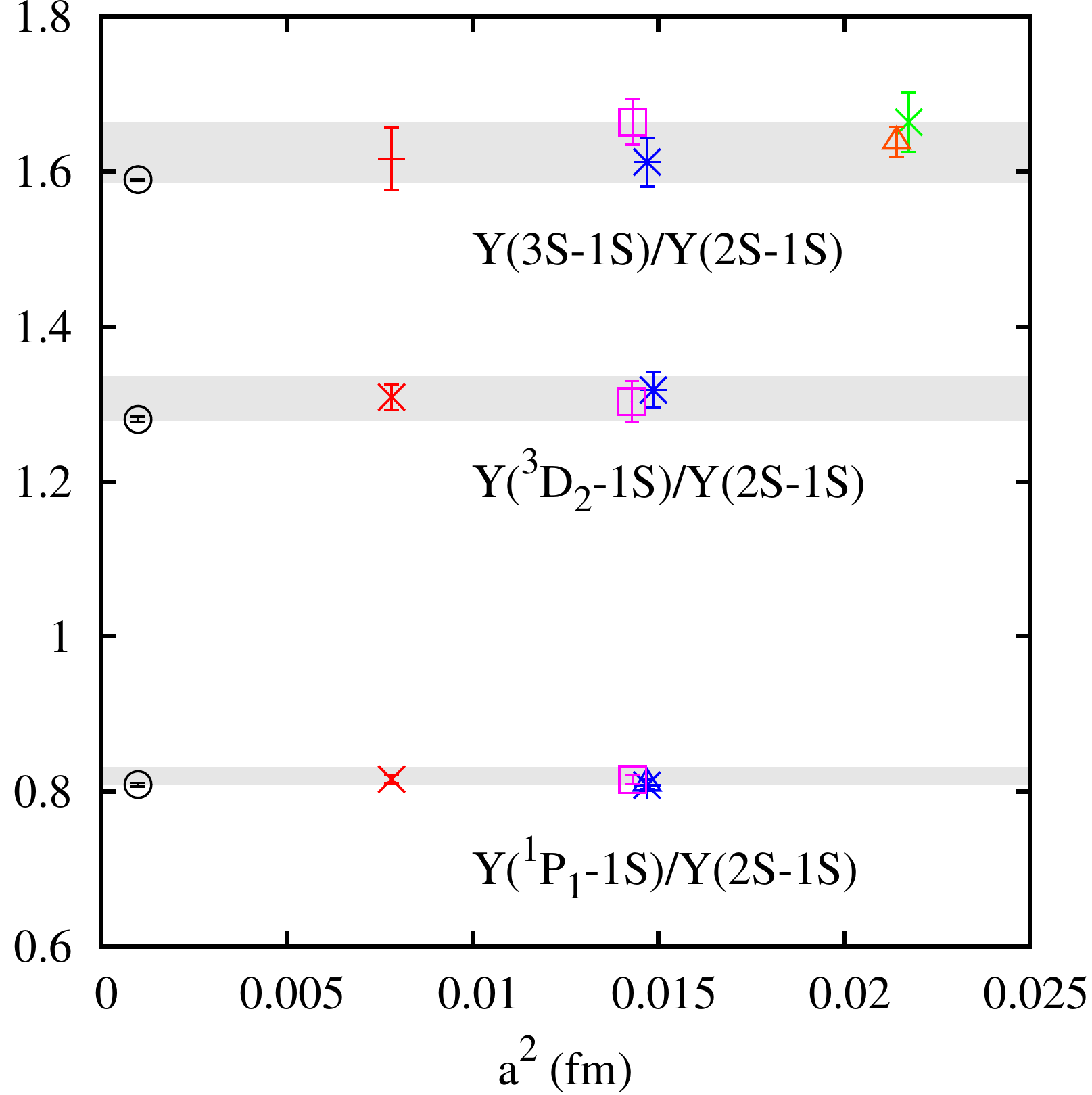}
\caption{ Results for the ratio of the $1^3D_2-1S$ 
splitting to the $2S-1S$ splitting in the $\Upsilon$
system plotted against the square of the lattice spacing 
determined from the $2S-1S$ splitting. 
Other ratios from~\cite{Dowdall:2011wh} are shown for comparison. 
The grey shaded bands give the physical result 
obtained from a fit to the data as described in the text, and 
with the full error of Table~\ref{tab:R_Derr}. 
The black open circles slightly offset from $a=0$ are from experiment~\cite{pdg}. 
 }
\label{fig:R_D}
\end{figure}

\begin{table}[t]
\caption{Fitted $D$-wave energies on each ensemble. Errors are from statistics/fitting only. $c_3 = 1.0$ on all ensembles, 
$c_4=1.25$ on sets 3 and 4 and 1.10 on set 5. 
$a\Delta(x) = aE(x)-aE(^3\overline{D})$. $R_X$ and $\Delta_X$ are defined 
in the text.
The $A_2$ irrep. on set 5 is fit separately and not included in the splittings.
}
\begin{ruledtabular}
\begin{tabular}{llll}
    & Set 3 	  & Set 4 	  	& Set 5 	\\
    & $am_b=2.66$ & $am_b=2.62$ 	& $am_b=1.91$ 	\\
\hline 
%%%%%%%%%%%%%%%%%%%%%%%%%%%%%%%%%%%%%%%%%%%%%%%%%%%%%%%%%%%%%%%%%%%%%%%%%%%%%%%%%%%
$aE(1^1D_{2E} )$   & 0.705(10)& 0.694(12) & 0.594(5)  \\
$aE(1^1D_{2T_2} )$ & 0.711(8) & 0.693(10) & 0.589(3)  \\
$aE(1^3D_{1T_1} )$ & 0.695(7) & 0.680(10) & 0.575(8)  \\
$aE(1^3D_{2E} )$   & 0.698(10)& 0.692(10) & 0.588(4)  \\
$aE(1^3D_{2T_2} )$ & 0.702(8) & 0.691(10) & 0.589(4)  \\
$aE(1^3D_{3A_2} )$ & 0.707(10)& 0.704(10) & 0.597(4)  \\
$aE(1^3D_{3T_1} )$ & 0.715(7) & 0.705(8)  & 0.596(4)  \\
$aE(1^3D_{3T_2} )$ & 0.714(7) & 0.696(9)  & 0.594(3)  \\
\hline
$a\Delta(1^1D_2)$  &  0.0029(31) &  0.0004(37) & 0.0027(27)  \\
$a\Delta(1^3D_1)$  & -0.0104(34) & -0.0137(44) & -0.0137(62) \\
$a\Delta(1^3D_2)$  & -0.0047(23) & -0.0021(21) & 0.0001(20)  \\
$a\Delta(1^3D_3)$  &  0.0078(22) &  0.0074(27) & 0.0069(20)  \\
\hline
$R_D $ 		   & 1.318(23)   & 1.303(26) &  1.309(16)  \\
\hline
%
%$a\Delta_{\bf L \cdot S} $ 	& -0.227(63)  & -0.237(80) &  -0.221(77) \\
%$a\Delta_{S_{ij}} $ 		& 0.007(16)   & -0.015(15) & -0.027(25)  \\
%$R_{\bf L \cdot S} $ 		& -2.18(64)   &  -2.44(85) &  -3.0(1.1)  \\
%$R_{S_{ij}} $ 			& -0.30(62)   &  0.63(60)  &  1.4(1.2)   \\
%
$a\Delta_{\bf L \cdot S} $ 	& 0.0038(11) & 0.0040(13) & 0.0037(13) \\
$a\Delta_{S_{ij}} $ 		& -0.0005(9) & 0.0009(9)  & 0.0016(15) \\
$R_{\bf L \cdot S} $ 		& 0.44(13)   & 0.49(17)	  & 0.60(21)   \\
$R_{S_{ij}} $ 			& -0.26(52)  & 0.53(50)   & 1.1(1.0) \\
\end{tabular}
\end{ruledtabular}
\label{tab:dval}
\end{table}

{\it Results.} The results from our fits for each $D$-wave lattice 
representation on each ensemble are 
given in Table~\ref{tab:dval}. 
We use $n_{\rm{exp}}=3$ on sets 3 and 4 and $n_{\rm{exp}}=4$ on set 5 
since these have the highest posterior probability~\cite{gplbayes}; values 
and errors have stabilised at this point and $\chi^2/{\rm{dof}} < 1$. 
We also give the ratio 
$R_D = (1^3D_2 - 1\overline{S})/(2^3S_1-1^3S_1)$ 
where $1\overline{S}$ is the spin-average of 
$\Upsilon$ and $\eta_b$ energies from~\cite{Dowdall:2011wh} and $1^3D_2$ is the 
dimension-weighted average of the lattice ${}^3D_{2E}$ and 
${}^3D_{2T_2}$ results.  

$R_D$ is plotted along with similarly defined 
$R_S$ and $R_P$ from~\cite{Dowdall:2011wh} 
in Figure~\ref{fig:R_D}. To obtain a physical result 
for $R_D$ we fit to the same form used in~\cite{Dowdall:2011wh} for 
$R_S$ and $R_P$, allowing for lattice spacing and sea quark 
mass dependence: 
\begin{eqnarray}
R &=& R_{\mathrm{phys}}[ 1 + 2b_l \delta x_l (1 + c_l(a\Lambda)^2) \\
&+&\sum_{j=1,2}c_j(a\Lambda)^{2j}(1+c_{jb}\delta x_m + c_{jbb}(\delta x_m)^2) ] .\nonumber 
\label{eq:fitxa}
\end{eqnarray}
Here $\delta x_l$ is $(am_l/am_s) - (m_l/m_s)_{\mathrm{phys}}$ 
for each ensemble. $(m_l/m_s)_{\mathrm{phys}}$ is 
taken from lattice QCD as 27.2(3)~\cite{Bazavov:2009bb}.  
$\delta x_m = (am_b -2.65)/1.5$ allows for $am_b$ effects 
from NRQCD in the discretisation errors over our range of 
$a$ values. 
$\Lambda$, taken as 500 MeV, sets the scale for physical $a$-dependence. 
Fit priors are as in~\cite{Dowdall:2011wh}: 1.0(0.5) 
on $R_{\mathrm{phys}}$; 0.0(0.3) on $a^2$ terms; 
0.0(1.0) on higher order in $a$; 0.0(0.015) on $b_l$.  
The physical result we obtain for $R_D$ is 1.307(30), 
after adding an additional NRQCD systematic error for 
missing $v^6$ terms~\cite{Dowdall:2011wh}. 
This is to be compared to the experimental value of 1.280(3). 
A complete error budget for $R_D$ is given in Table~\ref{tab:R_Derr}. 

\begin{table}
\caption{ Complete error budget for $R_D$ in \%. Finite volume 
and $m_b$ tuning errors are negligible. 
}
\label{tab:R_Derr}
\begin{ruledtabular}
\begin{tabular}{ll}
 & $R_D$  \\
\hline
stats/fitting		& 1.4  \\
$a$-dependence 		& 1.4  \\
$m_l$-dependence 	& 0.5  \\
NRQCD $am_b$-dependence & 0.1  \\
NRQCD systematics 	& 1.0  \\
electromagnetism/$\eta_b$ annihilation 	& 0.2  \\
\hline
Total & 2.3\%  \\
\end{tabular}
\end{ruledtabular}
\end{table}

\begin{figure}
%[t]
\includegraphics[width=0.9\hsize]{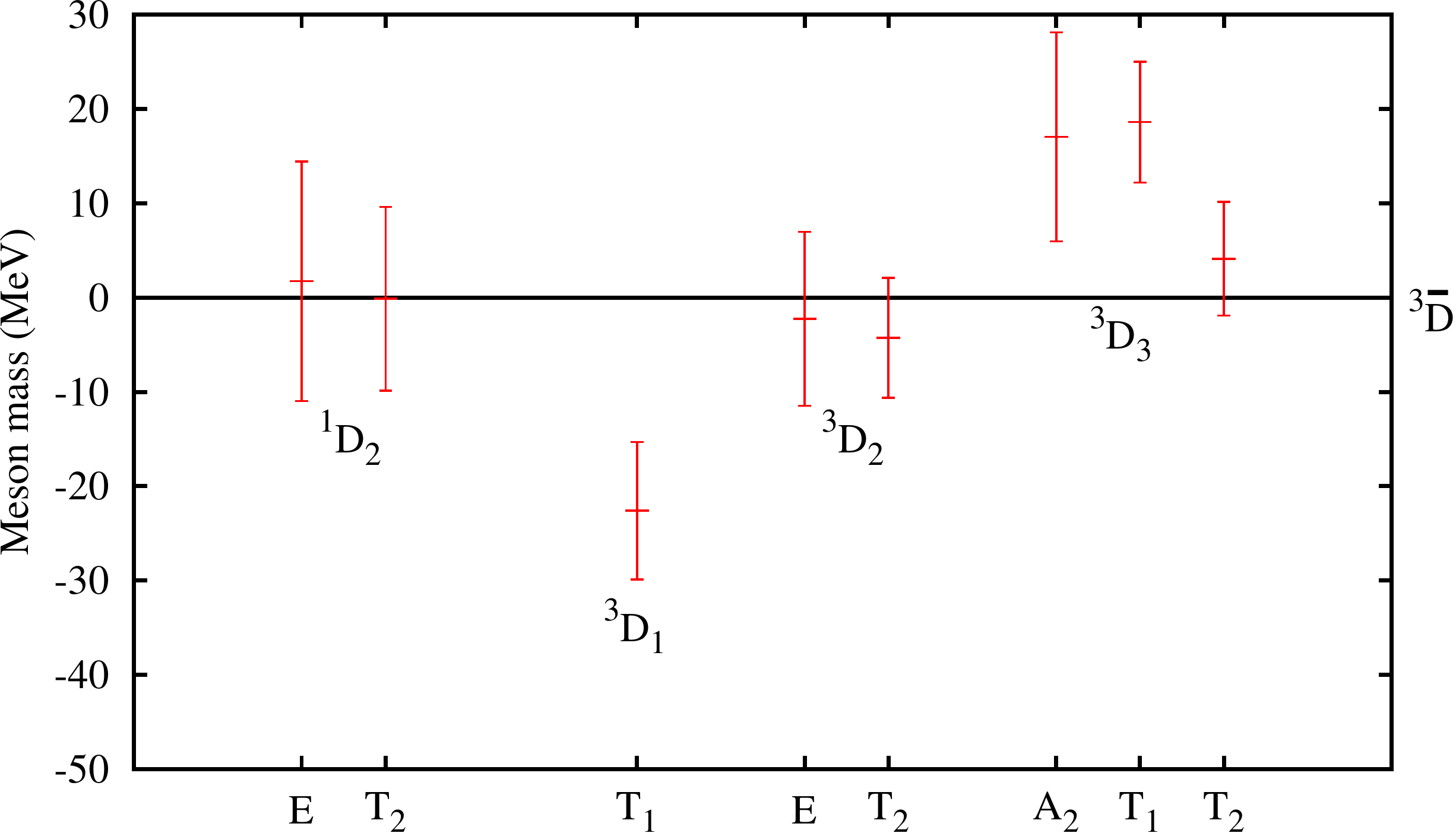}
\caption{Results for the separate irreducible 
representations of the lattice rotation group making up 
each continuum $D$-wave state on coarse set 4. }
\label{fig:coarse-allirreps}
\end{figure}
\begin{figure}
%[t]
\includegraphics[width=0.9\hsize]{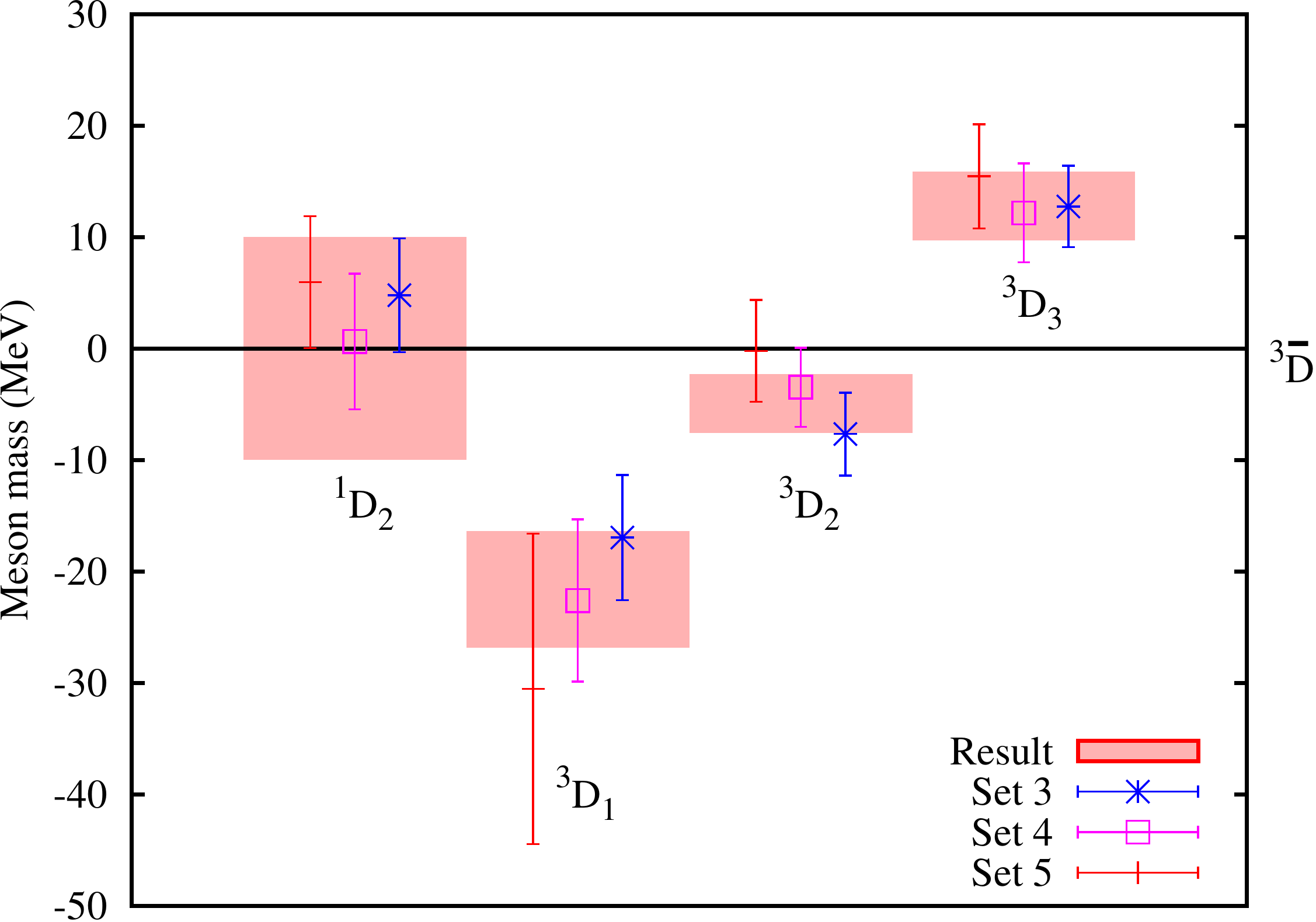}
\caption{$D$-wave masses plotted relative to the ${}^3D$ 
spin-average for all sets using 
the $2S-1S$ splitting to set the scale. The red shaded bands show our 
final results using ratios of combinations of splittings to 
those of $P$-wave states as described in the text. }
\label{fig:allensembles}
\end{figure}

\begin{figure}
%[t]
\includegraphics[width=0.9\hsize]{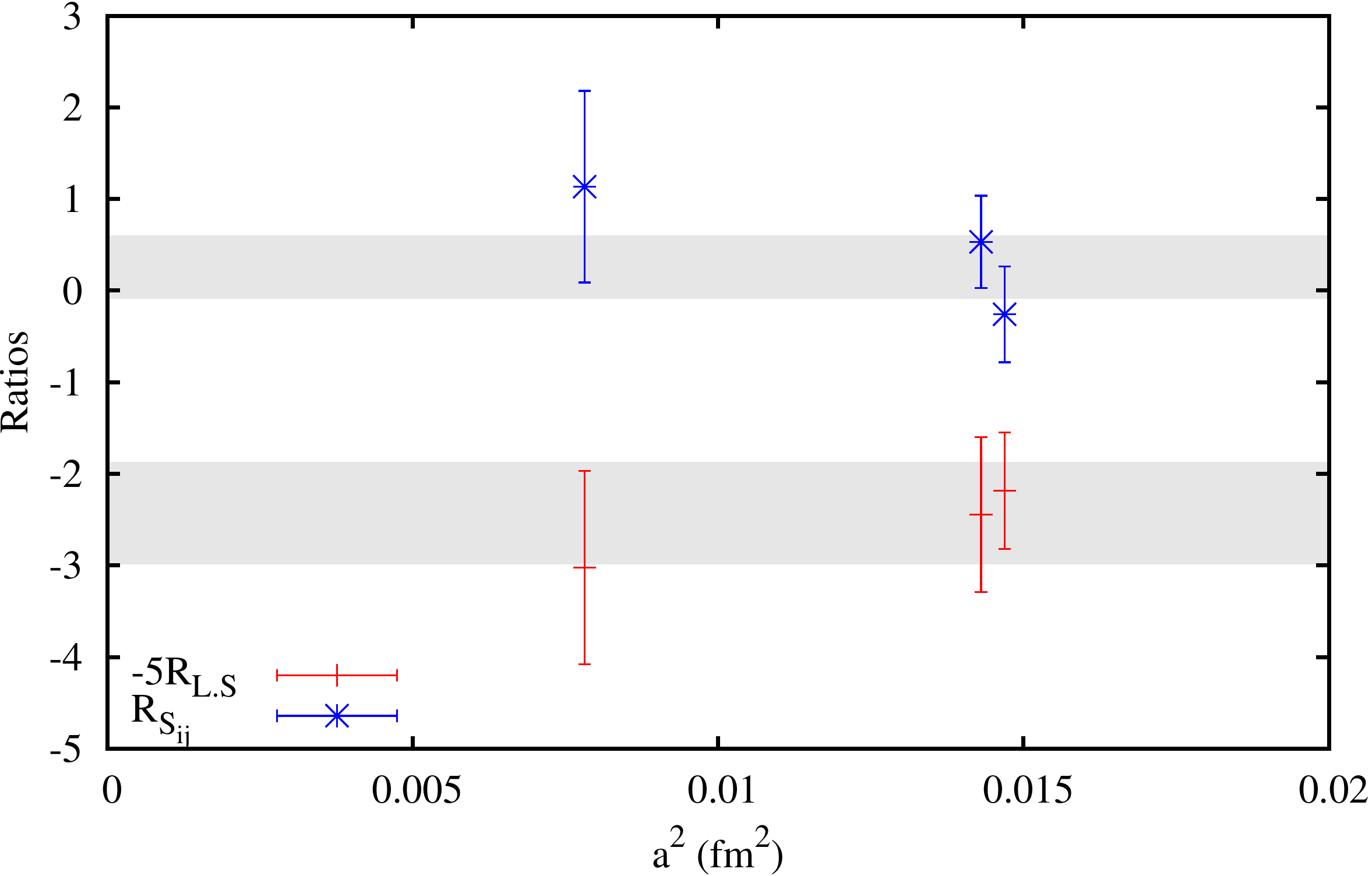}
\caption{Ratios $R_{S_{ij}}$ and $R_{\bf L\cdot S}$ (multiplied 
by -5 for clarity)  plotted 
against the square of the lattice spacing. The grey bands give 
our physical results.  }
\label{fig:ci-ratios}
\end{figure}

In Figure~\ref{fig:coarse-allirreps} we plot the 
masses of all 
the lattice representations relative to the 
spin average of all $1^3D$ states for coarse set 4, 
using the $2S-1S$ splitting to set the scale (Table~\ref{tab:params}). 
We see that the lattice representations for each 
spin agree well with each other within our sizeable 
statistical errors. The hyperfine splitting, between 
the ${}^1D_2$ and the spin average of ${}^3D$ states 
is expected to be very small, following results 
for $P$-wave states. We find it to be zero to within 10 MeV.   

Figure~\ref{fig:allensembles} shows the 
results from all three sets, 
using a dimension-weighted average of results, including 
the correlations from the fit, for 
the different lattice representations for the ${}^3D_3$ 
and ${}^{(1,3)}D_2$. Results are consistent between the 
fine and coarse sets and between different 
sea light quark masses for the two coarse sets. 

To arrive at a final result for $D$-wave
fine structure we study combinations of ${}^3D$ 
spin-splittings that are sensitive either to an $\bf{L}\cdot\bf{S}$
or to a tensor $S_{ij}$ interaction ($\bf{S}\cdot\bf{S}$ 
takes the same value for all ${}^3D$ states).
Writing 
\begin{equation}
M_J =\overline{M}({}^3D) + \Delta^D_{L\cdot S} \langle \bm{L}\cdot\bm{S} \rangle + \Delta^D_{S_{ij}} \langle S_{ij} \rangle
\end{equation}
gives 
\begin{eqnarray}
\Delta^D_{L\cdot S} &=& (14 M_3 - 5M_2 - 9M_1)/60 \nonumber \\
\Delta^D_{S_{ij}} &=& -7(2 M_3 - 5M_2 + 3M_1)/120 . 
\end{eqnarray}

Table~\ref{tab:dval} gives our results 
for these splittings. In Figure~\ref{fig:ci-ratios}
we plot ratios to the equivalent ${}^3P$ splitting combinations: 
$R_X = \Delta^D_X/\Delta^P_X$  with
$\Delta^P_{L \cdot S} = (5M_2-3M_1-2M_0)/12$ 
and $\Delta^P_{S_{ij}}=-5(M_2-3M_1+2M_0)/72$. 
Values for $\Delta^P$ for these ensembles 
are given in~\cite{Dowdall:2011wh} (without factors 
of 1/12 and -5/72).
The experimental values are $\Delta^P_{L \cdot S}$ = 13.65(27) MeV and 
$\Delta^P_{S_{ij}}$ = 3.29(9) MeV~\cite{pdg}. 
The advantage of using these combinations is that they depend 
purely on one of the spin-dependent coefficients of the NRQCD 
action. On set 5 we did not use exactly the same values for $c_3$ 
and $c_4$ in our study of $P$ and $D$ waves. However we can 
correct for this in Figure~\ref{fig:ci-ratios} 
since $\Delta_{S_{ij}} \propto c_4^2$ and 
$\Delta_{L\cdot S} \propto c_3$. Once this slight adjustment 
is done the dependence on $c_{3,4}$ cancels 
between $P$ and $D$ states and 
so errors from the uncertainty in these coefficients are much reduced.  

We fit the fine-structure $R$ values to the same form 
used earlier in eq.~\ref{eq:fitxa} 
to extract physical results:
\begin{equation}
R_{L \cdot S} = 0.49(11); \quad R_{S_{ij}} = 0.26(35). 
\end{equation}
We have included an additional systematic error of 10\% to allow for 
missing $v^6$ terms from our NRQCD action but the 
lattice statistical error dominates. We then combine 
the $R$ values with experimental results from $1P$ 
levels to give the following ${}^3D$ splittings: 
\begin{eqnarray}
{}^3D_3 - {}^3D_1 &=& 34(8) \mathrm{MeV} \nonumber \\
{}^3D_3 - {}^3D_2 &=& 18(5) \mathrm{MeV} \nonumber \\
{}^3D_2 - {}^3D_1 &=& 17(6) \mathrm{MeV} .
\end{eqnarray}

Our fine structure splittings are somewhat larger than 
typical results from potential 
models~\cite{Kwong:1988ae, Brambilla:2004wf}, where 
the ${}^3D_3-{}^3D_1$ splitting lies in the range 10-20 MeV. 
This can be traced to a larger value for $R_{L \cdot S}$ 
than is obtained, for example, in~\cite{Kwong:1988ae}, 
based on specific forms for the spin-dependent potentials. 

\begin{figure}
%[t]
\includegraphics[width=0.8\hsize]{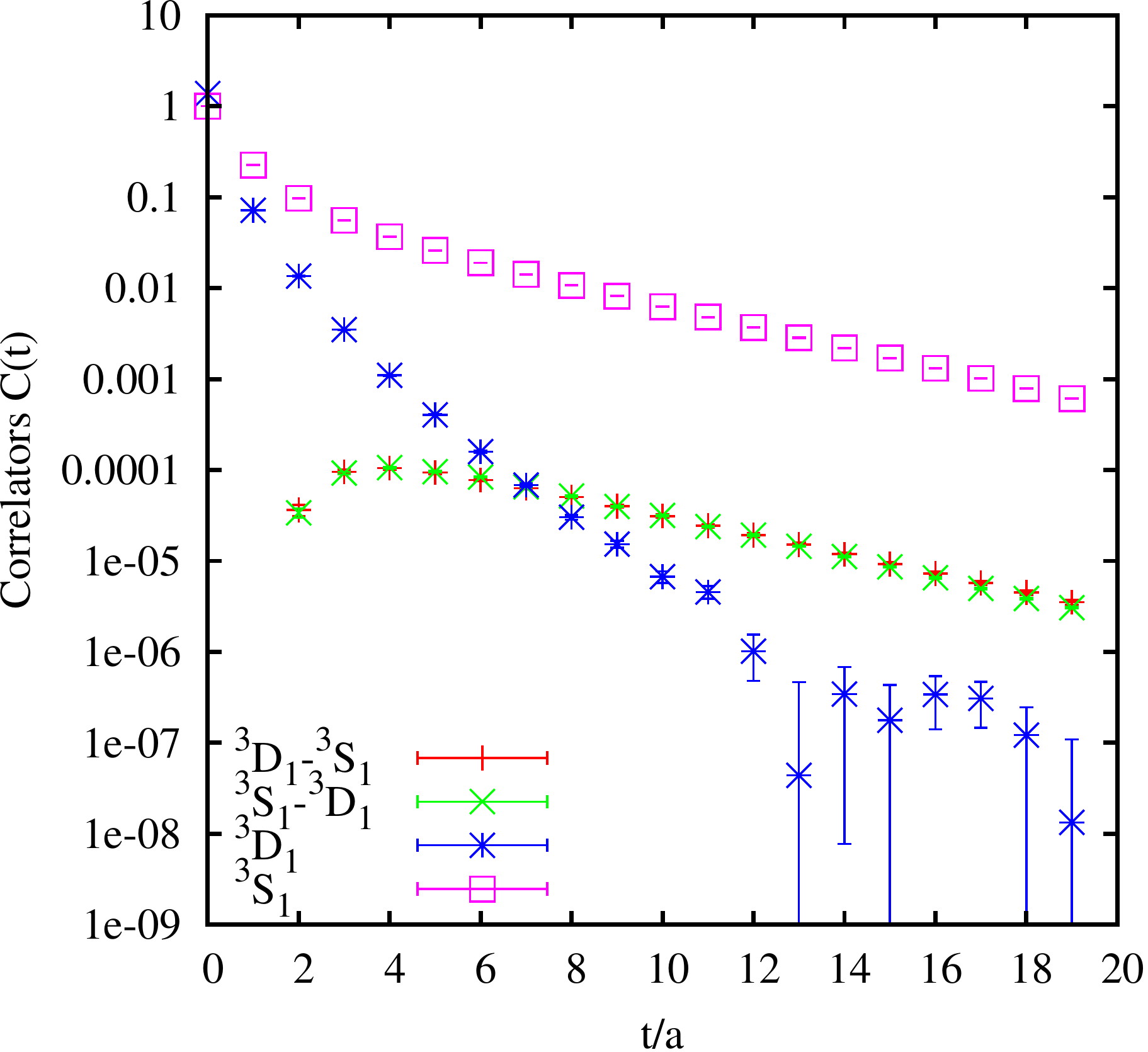}
\caption{Correlators made from different combinations of 
local ${}^3S_1$ and ${}^3D_1$ 
operators at source and sink, plotted with a logarithmic $y$ axis as 
a function of lattice time, $t$. Results are from set 3. }
\label{fig:s-d-wave}
\end{figure}

One issue that we have neglected above is that the ${}^3D_1$ 
state has $J^{PC}=1^{--}$ in common with ${}^3S_1$ states. On 
the lattice, in principle, any operator with $1^{--}$ quantum 
numbers will be able to create all $1^{--}$ states. In practice 
the amplitude for ${}^3S_1$ states to be created by the operators 
that we use for the ${}^3D_1$ is very small and vice versa. We 
illustrate that in Fig.~\ref{fig:s-d-wave} where we show correlators from set 3 that 
use a local ${}^3S_1$ or ${}^3D_1$ operator at source and sink 
compared to the cross-correlator that has a 
local ${}^3S_1$ operator at the source 
and ${}^3D_1$ at sink or vice versa. The cross-correlator is 
much smaller in magnitude than either of the diagonal correlators 
at small $t$ values. The exponential fall-off (as seen in the 
slope of the log plot) of the cross-correlator matches that 
of the ${}^3S_1$ 
correlator at large times, where the ${}^3D_1$ correlator fall-off
is dominated by that of the heavier ${}^3D_1$ state. If we fit
the complete set of ${}^3S_1$ and ${}^3D_1$ correlators together, 
including the local cross-correlators of Fig.~\ref{fig:s-d-wave}, we obtain 
results in agreement with our separate fits for ${}^3S_1$ (in~\cite{Dowdall:2011wh}) and 
${}^3D_1$ masses. We also find, for example, that the amplitude 
$a({}^3D_{1,local}, \Upsilon)$ from eq.~\ref{eq:fit} is 0.0052(1) times 
that of $a({}^3S_{1,local}, \Upsilon)$.    

%%%%%%%%%%%%%%%%%%%%%%%%%%%%%%%%%%%%%%%%%%%%%%%%%%%%%%%%%%%%%%%%%

{\it Conclusions.} We give the first full lattice QCD 
results for the $D$-wave states of bottomonium including 
the fine structure. We obtain a mass of 10.179(17) GeV for 
the $1^3D_2$ to 
be compared with 10.1637(14) GeV from experiment~\cite{pdg}. 
Using the experimental 
result for the $1^3D_2$ mass we predict masses of 
10.181(5) GeV for the $1^3D_3$, 
10.147(6) GeV for the $1^3D_1$ and 10.169(10) for the $^1D_2$.  

{\it Acknowledgements.} We are grateful to MILC for the use of their 
gauge configurations and to Peter Lepage for comments. 
We used the Darwin Supercomputer 
under the DiRAC facility jointly
funded by STFC, BIS 
and the Universities of Cambridge and Glasgow. 
This work was funded by STFC and the EU Erasmus programme.

\bibliography{ups}

\end{document}